\documentclass[prl,twocolumn,showpacs,lengthcheck,amsmath,amssymb]{revtex4}
\usepackage{graphicx}
\usepackage{epstopdf}
\usepackage{enumerate}
\usepackage{braket}
\usepackage{color} 
\newcommand \beq{\begin{eqnarray}}
\newcommand \eeq{\end{eqnarray}}

\begin{document}

\title{Antiferrosmectic ground state of two-component dipolar Fermi gases \\ 
-- an analog of meson condensation in
nuclear matter}
\author{Kenji Maeda,$^{1}$ Tetsuo Hatsuda,$^{2,3}$ and Gordon Baym$^{4}$}
\affiliation{
$^{1}$Department of Physics, Colorado School of Mines, Golden, CO 80401, USA\\
$^{2}$Department of Physics, The University of Tokyo, Tokyo 113-0033, Japan \\
$^{3}$Theoretical Research Division, Nishina Center, RIKEN, Wako 351-0198, Japan \\
$^{4}$Department of Physics, University of Illinois, 1110 W. Green Street,
Urbana, Illinois 61801, USA\\  \\
}

\begin{abstract}
We show that an antiferrosmectic-C phase has lower energy at high densities 
than the non-magnetized Fermi gas and ferronematic phases 
in an ultracold gas of fermionic atoms, or molecules, with large magnetic dipole moments. 
This phase, which  is analogous to
 meson condensation in dense nuclear matter, is a one-dimensional periodic structure 
in which the fermions localize in layers with their pseudospin direction aligned parallel to the layers, 
and staggered layer by layer.   \end{abstract}
\maketitle

Ultracold atoms with strong dipole-dipole interactions
offer a unique opportunity to study the properties of many-body systems 
with long-range anisotropic interactions \cite{Baranov:2008}. 
Such systems enable one to realize in the laboratory analogs of meson condensation in nuclear matter,
as a consequence of the similarities of the electric and magnetic dipole interactions to the nuclear tensor force.
Systems of magnetic  atoms  such as  dysprosium --
which has the largest magnetic moment of all stable atoms, some 10 times that of alkali atoms, 
and currently trapped and cooled by Lev's group \cite{Lev:2010,Lev:2012} --
can exhibit novel quantum dipolar phases \cite{huidy}. 
In particular, clouds of the fermionic isotopes, $^{161}$Dy and $^{163}$Dy 
may be used to reveal different aspects of the dipolar Fermi gas. 
Examples are:
spheroidal deformations of the Fermi surface in one-component Fermi systems with dipolar interactions, 
studied by Sogo et al. \cite{Sogo:2009}; and
ferronematic  order with a deformed Fermi surface in two-component dipolar Fermi systems, 
pointed out by  Fregoso et al. \cite{Fradkin:2009,Fradkin2}, which may appear when 
the repulsive contact interaction or the dipole-dipole interaction is sufficiently large. 
On the other hand, two-component uniform Fermi gases become unstable
against long-wavelength local magnetization as the interactions become strong; 
see, e.g., RPA analyses in Refs.~\cite{Yamaguchi:2010,Sogo:2012}. 
The ground state structure with local magnetization and its competition with the ferronematic state 
in the  strong coupling region have not been explored so far. 

In this paper, we propose a variational state of two-component quantum dipolar Fermi systems,
with spatially varying magnetization, motivated by studies of classical dipoles on a three dimensional cubic lattice 
\cite{Luttinger:1946} 
and of meson-condensed systems in nuclear physics 
\cite{Kunihiro:1978,Matsui:1978,Baym:1979,Tamagaki:1993}. 

\begin{figure}[t]
\begin{center}
\includegraphics[width=8.5cm]{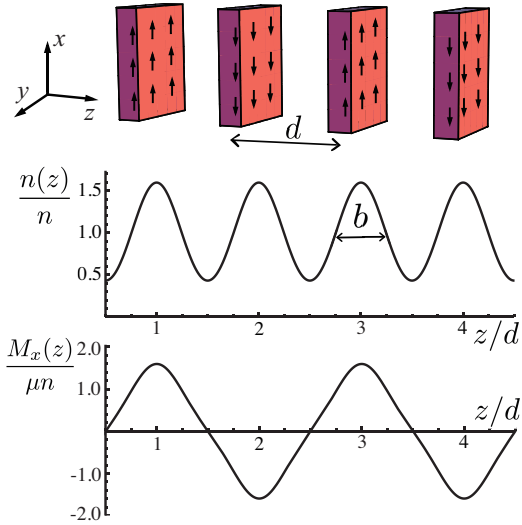}
\end{center}
\vspace{-0.5cm}
\caption{
The uppermost panel illustrates the magnetization profile, 
and the lower two panels the spatial distribution along $z$-direction of the normalized number density and magnetization, 
Eqs.~(\ref{AFSC:spin}), in the AFSC state for $d = \sqrt8 b$. 
}
\label{Fig1}
\end{figure}

The most advantageous structure of a dipolar system is determined by competition among 
the dipolar interaction favoring magnetization (spin polarization) varying in direction in space, 
short range repulsions favoring aligned spins, 
and particle kinetic energies favoring spatially uniform systems.  
Our new state is an {\it antiferrosmectic-C} (AFSC) phase \cite{chaikin}, 
a one-dimensional layered structure with alternating dipoles parallel to the layers.
The AFSC structure is illustrated in the top of Fig.~\ref{Fig1} -- layered in the $z$-direction 
with a staggered non-sinusoidal magnetization in the $x$-direction: 
$\braket{\vec{M}(\vec{r}\,)}= M_x(z) \hat{x}$ \cite{note1}.  
As we show, this state is energetically more favorable than the ferronematic phase 
over a wide region of dipole and short range repulsive interaction strengths. 
The AFSC phase has no net magnetization, but utilizes the dipole-dipole attraction efficiently.

We model the system of ultracold fermionic atoms or molecules
 with dipole-dipole interactions and short range repulsion 
 in terms of the Hamiltonian: 
\beq
\label{eq:dipolar-hamiltonian}
H&=&
\int\!{\rm d}\vec{r}_1\:
\frac{\nabla\Psi^{\dagger}(\vec{r}_1\,)\cdot\nabla\Psi(\vec{r}_1\,)}{2m}
\nonumber\\
&+&
\frac12 
\sum_{i,j=1}^3\sum_{\alpha,\alpha' ,\, \beta,\, \beta'}^{\uparrow,\downarrow}
\int{\rm d}\vec{r}_1{\rm d}\vec{r}_2\: 
\psi^{\dagger}_{\alpha}(\vec{r}_1)\psi^{\dagger}_{\beta}(\vec{r}_2)
\qquad\quad
\\
&&
\qquad\qquad\quad
\times
V(\vec{r}_1,\vec{r}_2)_{\alpha\alpha',\, \beta\beta'}^{ij}\psi_{\beta'}(\vec{r}_2)\psi_{\alpha'}(\vec{r}_1)
~.
 \nonumber
\eeq
Here $\Psi=(\psi_{\uparrow}, \psi_{\downarrow})$ describes 
fermions of mass $m$ in two hyperfine states with a transition magnetic moment $\mu$ \cite{note2}. 
For simplicity we refer to this internal degree of freedom as ``spin." 
The $\vec{\sigma}$ are the Pauli spin matrices. 
We take $\hbar = 1$ throughout. 
Our model is characterized by the interaction potential 
\beq
V(\vec{r}_1,\vec{r}_2)_{\alpha\alpha',\, \beta\beta'}^{ij}
&=&
\frac{\mu^2}{r^3}
\bigl\{
\sigma^i_{\alpha\alpha'}(\delta_{ij}-3\hat{r}_i\hat{r}_j)\sigma^j_{\beta\beta'}
\bigl\}
\nonumber\\
&&
+\: g \delta_{\alpha\alpha'}\frac{\delta_{ij}}{3}\delta(\vec{r}_1-\vec{r}_2)\delta_{\beta\beta'}~, 
\label{Vpot}
\eeq
where $r=|\vec{r}_1-\vec{r}_2|$ and $\hat{r}_i=(\vec{r}_1-\vec{r}_2)_i/r$. 
The first term is the long-range dipole-dipole interaction potential. 
The second term is the short-range contact interaction potential, which 
operates only between different species owing to the Pauli principle.
We assume for simplicity that  it is always repulsive, i.e., $g>0$. 
 We define the dimensionless coupling constants, 
\beq
  \lambda_d~=~ n \mu^2/\epsilon_{_F}, \quad   \lambda_s~=~ n g/\epsilon_{_F},
\eeq
where $n$ is the average fermion density and 
$\epsilon_{_F}$ is the Fermi energy of the two-component fermions at the density. 
A coupled channel RPA analysis of the spin-triplet ($S = 1$) excitation 
with angular momentum $L=0$ and $L = 2$ states \cite{Sogo:2012} shows that 
the non-magnetized Fermi gas becomes unstable  against a spatially varying magnetization 
 beyond the critical line in the $\lambda_s$-$\lambda_d$ plane:
\beq
\label{eq:sogo}
\left(1-\frac34 \lambda_s -2\pi \lambda_d\right)
\left(1+\frac{\pi}{2}\right)
-\frac{\pi^2}{2}\lambda_d^2
&=&0~. 
\eeq

To seek the ground state of dipolar fermions in the strong coupling
region beyond this critical line,  
we introduce a set of normalized basis states for particles localized in the $z$-direction, 
and in plane wave states in the transverse direction, 
\beq
\phi_{\ell,\vec q_\perp }(x,y,z)
&=&
\frac{e^{-(z-d \ell )^2/2b^2}}{(\pi b^2)^{1/4}}
\frac{e^{i\vec q_\perp \cdot  \vec r_\perp}}{\sqrt{V_\perp}}  \chi_\ell ,
\label{eq:AFSC}   
\eeq
where the integer $\ell$, ranging from $-\infty$ to $+\infty$ labels the layer, 
$\vec q_\perp$ is the transverse momentum, $\vec r_\perp = (x,y)$, $V_\perp$ is the transverse two-dimensional volume, 
$d$ is the distance between neighboring layers, $b$ is the width of the layer, 
and the $\chi_\ell =(1,(-1)^{\ell+1})/\sqrt2$ are staggered spinors, with the center site, $\ell = 0$, spin-down with respect to $\sigma_x$. 
We assume that the fermions are well localized in the $z$-direction, with $d\gtrsim \sqrt2 b$, 
and in the AFSC ground state fill these states with all $\ell$, and $q_\perp$ up to the transverse Fermi momentum, 
$q_{F\perp} = \sqrt{4\pi nd}$.

The expectation values of the number density, $\Psi^{\dagger}\Psi$, 
and local magnetization, $\mu \Psi^{\dagger}\vec \sigma \Psi$, in the AFSC state are 
\beq
\langle n(\vec{r}\,) \rangle
&=&
\frac{nd}{b\sqrt{\pi}}\sum_{\ell =-\infty}^\infty e^{-(z-d \ell)^2/b^2}, 
\label{AFSC:number} 
\nonumber \\
\langle {M_x}(\vec{r}\,) \rangle
&=& 
-\frac{\mu nd}{b\sqrt{\pi}}
\sum_{\ell =-\infty}^\infty(-1)^\ell  e^{-(z-d \ell)^2/b^2}, 
\label{AFSC:spin}
\eeq
and 
$\langle {M_y}(\vec{r}\,) \rangle=\langle {M_z}(\vec{r}\,) \rangle=0$. 
The AFSC state is indeed a layered structure in density, with staggered local magnetization, 
as illustrated in the middle and lower panels of Fig.~\ref{Fig1}. 
Using the Poisson summation formula, we can write the density and magnetization as 
\beq
\langle n(\vec{r}\,) \rangle
&=& 
n+2n\sum_{j=1}^\infty e^{-j^2 \pi^2/\Gamma}\cos\left(2 j \pi  z/d\right),
\label{n_mode_exp}\\
\langle {M_x}(\vec{r}\,) \rangle
&
\! \! \! \! \! 
=&
\! \! \! \! \! 
 - 2\mu n \sum_{j=1}^\infty  e^{-(2j-1)^2 \pi^2\!/4\Gamma}\!
\cos\left\{(2j\!-\!1)\pi z/d\right\},\nonumber
\label{m_mode_exp}
\eeq
where the dimensionless localization parameter $\Gamma =(d/b)^2$ 
is the square of the ratio of the layer distance and the Gaussian width.

\begin{figure}[t]
\begin{center}
\includegraphics[width=8.5cm]{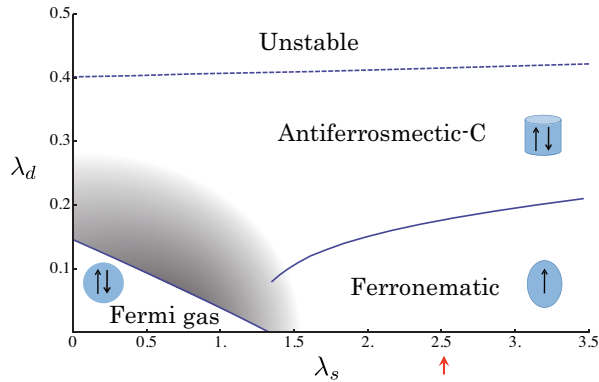}
\end{center}
\vspace{-2.0cm}
\caption{
Schematic phase structure of dipolar fermions as a function of $\lambda_s$ and $\lambda_d$, 
showing the non-magnetized Fermi gas phase, the locally weakly magnetized state shown by the shaded area, 
the ferronematic region, and the antiferrosmectic-C region. 
Beyond the upper dashed line the system becomes unstable against collapse.}
\label{Fig2}
\end{figure}

The energy density of the AFSC phase, ${\cal E}(\Gamma,\! \alpha)$, with $\alpha= 1/8\pi n d b^2$, 
in units of the energy density of the free Fermi gas is 
\beq
\frac{{\cal E}(\Gamma,\! \alpha)}{\frac35 n \epsilon_F}
\!&=&\!
\frac{10}{3(3\pi)^{2/3}}\Gamma^{1/3}\alpha^{2/3}
+\frac{5}{3(3\pi)^{2/3}}\Gamma^{1/3}\alpha^{-1/3}
\nonumber \\
 & &  -
\frac{20\pi}{3}\lambda_d
\sum_{j=1}^\infty 
e^{-(2j-1)^2\pi^2\!/2\Gamma}
\left\{\frac13-F(\alpha)\right\} \nonumber \\
\! \! & &  \! \! + \frac{5}{6}
\lambda_s
\biggl\{
\frac12
\!-\!\sum_{j=1}^\infty 
\bigl[
e^{-(2j-1)^2\pi^2\!/2\Gamma}
-e^{-2j^2\pi^2\!/\Gamma}
 \bigl]\!
\biggl\}. \nonumber\\
\label{AFSC:energy}
\eeq
The first term on the right is the one-dimensional  zero-point energy in the $z$-direction, 
and the second term is the two-dimensional kinetic, or Fermi, energy within a layer. 
We note that $2\alpha$ is the dimensionless ratio of these two terms. 
The third term arises from the dipole-dipole interaction, 
with $1/3$ and $-F(\alpha)$ the direct and exchange contributions, respectively, where
\beq
\lefteqn{F(\alpha)}
\\
\!\!&=&\!\!
 \alpha\! \int_0^{\infty}\!{\rm d}s J_1^2 \!
\left(\!\sqrt{2s/\alpha}\right)
 e^{s} 
\biggl\{
 \frac{2s+1}{s} K_0(s)
 \!-\! 2K_1(s)
 \biggl\},
 \nonumber
\eeq
and $J_{i}$  and $K_{i}$  are the Bessel functions of the first kind and second kind, respectively. 
The final, positive, term in Eq.~(\ref{AFSC:energy}) is the contribution of the contact interaction. 
The lowest energy AFSC state is obtained by minimizing ${\cal E}(\Gamma,\! \alpha)$ 
with respect to $\Gamma$ and $\alpha$ for given $\lambda_d$ and $\lambda_s$. 
It is sufficient to  take only the $j=1$ mode in Eq.~(\ref{AFSC:energy}) (one-mode approximation) 
to obtain the ground state energy to a few percent  accuracy for $\Gamma \lesssim 10$, 
the regime in which particles are not too well localized in the layers. 
In constructing the phase diagram of the system as functions of 
the short range interaction, measured by $\lambda_s$, 
and the dipole-dipole interaction, measured by $\lambda_d$, 
as shown in Fig.~\ref{Fig2}, 
we compare the minimum ${\cal E}(\Gamma,\! \alpha)$ 
with the energy of the interacting Fermi gas phase 
[$\epsilon_{\rm FG}
={\frac35 n \epsilon_F}\left( 1+\frac{5}{12}\lambda_s \right) $] 
and with that of the fully polarized ferronematic state \cite{Fradkin:2009,note3}.

The resultant phase diagram, shown schematically in Fig.~\ref{Fig2},
is composed of four regions:
(I) the non-magnetized Fermi gas phase (the lower left corner), which has a spherical Fermi surface with equal population of both species,
(II) the AFSC phase  (the upper right region) which has a cylindrical Fermi surface with equal populations of both species,
(III) the ferronematic phase (the lower right region) which has a spheroidal Fermi surface with only a single species, and
(IV) an unstable region. 
Determining the detailed structure of the intermediate locally weakly magnetized state, 
shown as the shaded area~\cite{Sogo:2012,Matsui:1978}, and the question of whether this phase evolves 
smoothly into the AFSC phase with increasing dipole interaction strength, are interesting open problems beyond the scope of this paper.

\begin{figure}[t]
\begin{center}
\includegraphics[width=8cm]{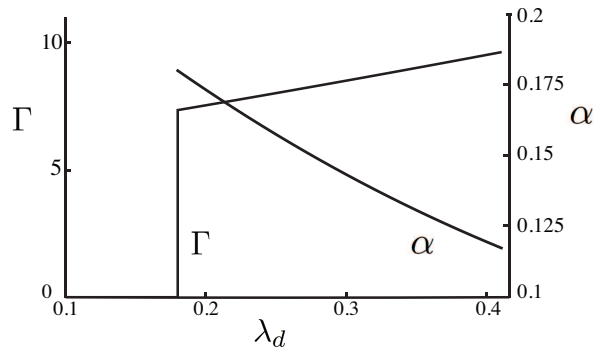}
\end{center}
\vspace{-0.5cm}
\caption{
Optimal values of $\Gamma$ and $\alpha$ as a function of $\lambda_d$ 
for $\lambda_s=2.5$. 
}
\label{Fig3}
\end{figure}

 In Fig.~\ref{Fig3} we plot the localization parameter $\Gamma$ and the kinetic energy ratio $\alpha$ vs. 
$\lambda_d$ for $\lambda_s=2.5$, corresponding to the red arrow in Fig.~\ref{Fig2}. 
The exchange contribution of the dipole-dipole interaction in the AFSC state 
is of order 30\% of the direct term, decreasing with increasing $\lambda_d$.

The different regions of the phase diagram are related as follows: 

\noindent {\it From Fermi gas to AFSC}: 
The energy of the non-magnetized Fermi gas is increased only by the
contact interaction, not by the dipolar interaction. 
On the other hand, the AFSC phase has a net decrease in energy from the dipolar attraction 
between neighboring layers, despite the increase of the fermion kinetic energy 
due to the spatial localization in the $z$-direction. 
Therefore, as $\lambda_d$ increases, the non-magnetized  Fermi gas, after making a transition to 
a locally weakly magnetized phase (the shaded region in Fig.~\ref{Fig2}), passes into the AFSC phase. 

\noindent{\it From Fermi gas to ferronematic}: 
The fully polarized ferronematic phase has only one component, 
and it does not experience the contact repulsion, 
although its Fermi energy increases in comparison to the two-component Fermi gas. 
Therefore, there is a transition from the Fermi gas phase to the ferronematic phase 
for sufficiently large $\lambda_s$, as long as $\lambda_d$ is small. 

\noindent{\it From ferronematic to AFSC}: 
In the fully polarized ferronematic phase with small $\lambda_d$, the Fermi surface is deformed from spherical
due to the anisotropy of the dipole interaction \cite{Sogo:2009,Fradkin:2009}. 
As $\lambda_d$ increases, however, the one-dimensional periodic structure emerges spontaneously, 
enhancing the dipolar attraction in a two-component system. 
This leads to a transition from weak to strong deformation of the Fermi surface,
replacing the ferronematic phase by the AFSC phase for a wide range of $\lambda_d$. 

\noindent{\it Instability}: 
For $\lambda_d$ above the dashed line in Fig.~\ref{Fig3} ($\lambda_d \simeq 0.40\sim 0.43$), 
the AFSC phase becomes unstable due to large attraction between the layers. 
The dashed line is determined by the compressibility, 
$K^{-1} = n\partial P/\partial n = n^2 \partial^2{\cal E}/\partial n^2$, vanishing. 
A similar instability of the ferronematic phase has been reported in \cite{Sogo:2009,Fradkin:2009}. 
Determining the stable states of the system above the dashed line remains a problem for future work.

We now relate the present model to laboratory systems. 
Numerically, the coupling constants are
$\lambda_d = 2.7\times 10^{-3} m_{100}n_{12}^{1/3}\mu_{10}^2$ 
and
$\lambda_s=2.6\times10^{-2} n_{12}^{1/3} a_{10}$, 
where $m_{100}$ is the atomic (or molecular) mass in units of 100 proton masses, 
$n_{12}$ is the density in units of $10^{12} {\rm cm}^{-3}$, 
$\mu_{10}$ is the magnetic dipole moment in units of ${10 \mu_B}$, 
and $a_{10}$ is the effective scattering length, $g m /4\pi$, in units of  10 nm. 
Then  $\lambda_d/\lambda_s = 0.1\times m_{100}\mu_{10}^2 /  a_{10}$. 
The distance $d$ between layers in the AFSC phase is $\sim 10^2  n_{12}^{-1/3}$ in $\mu$m, 
which may be fine compared to the size of ultracold atomic gases. 
To reach into the AFSC phase with Dy atoms 
would require $n\sim 5\times10^{17} $ cm$^{-3}$, a density which can be greatly reduced by resonantly enhancing $a$. 
Molecules with larger masses and magnetic moments, 
e.g. diatomic molecules of Dy,  
have larger $\lambda_d$ 
 also enabling the AFSC phase to be realized at smaller density. 

For dipoles with more than two internal states, 
the  relation of diagonal and off-diagonal moments is modified, so that in Eq.~(\ref{Vpot}) 
$\sigma_z$ must be replaced by diag$\{A,B\}$ where 
$A = \bra{\uparrow}\mu_z\ket{\uparrow} /\mu$, 
and $B =  \bra{\downarrow}\mu_z\ket{\downarrow}/\mu$, 
with $\mu_z$ the $z$-component of the moment operator, and $\mu$ the transition moment. 
This correction enhances local magnetic fields along the $z$-direction, 
which for a system uniform in the $x$- and $y$-directions 
can at most be constant (and generally small compared with trapping fields).

Calculation of the finite temperature phase diagram remains an open problem. 
The scale of energies in the various phases at zero temperature are of order the Fermi energy, 
and so one expects significant structure to emerge at temperatures below the Fermi temperature. 
The regime of stability at finite temperature of the two-dimensional sheets in the AFSC phase in finite volume, 
particularly against a Landau-Peierls instability, needs to be determined.
Finally, one should consider other possible phases in the system, e.g., 
AFSC states with population imbalance, states with structure in the sheets such as a ``spaghetti phase" \cite{benjamin},
and superfluid states of spin-triplet p-wave paired atoms which could emerge at large $\lambda_d$.

   As Luttinger and Tisza \cite{Luttinger:1946} showed,  
 the lowest energy configuration of classical dipoles 
 on a cubic lattice has the same dipole configuration as the present AFSC phase.    
 Thus we can reasonably expect that the AFSC state can be realized in an optical lattice 
 and might even be enhanced when the pitch of the spontaneous localization 
 is commensurate with that of the lattice potential.

 The dipole-dipole interaction is caused by exchange of the vector potential $\vec{A}(\vec{r})$ 
between dipolar atoms. 
Therefore spatially varying magnetizations $\langle \vec{M} (\vec{r})\rangle$ in the AFSC phase
correspond to the spontaneous formation of a non-zero $\langle \vec{A}(\vec{r}) \rangle$,
through the Maxwell equation: $\nabla^2 \langle \vec{A}(\vec{r}) \rangle
= - 4 \pi \nabla \times \langle \vec{M} (\vec{r})\rangle$.
Very similar {\it meson condensed states have been proposed} in  nuclear physics 
where the $\rho$-meson
 in nuclear matter 
corresponds to $\vec{A}(\vec{r})$ for magnetic dipolar atoms, and 
 the $\pi$-meso corresponds to 
the scalar potential $A_0(\vec{r})$ for electric dipolar atoms.
Such a correspondence may open the interesting future possibility
of learning properties of states of dense matter inside neutron stars via cold atom experiments in the laboratory.
    
\begin{acknowledgments}
We thank Benjamin Lev, Eduardo Fradkin, and Benjamin Fregoso for helpful discussions, 
and the Aspen Center for Physics, supported in part by NSF Grant PHY10-66293, where this work originated. 
This research was supported in part by NSF Grant PHY09-69790 and 
JSPS Grants-in-Aid for Scientific Research Nos.18540253, 22340052.
K.M. was supported by JSPS Research Fellowship for young scientists and AFOSR. 
In addition G.B. wishes to thank the G-COE program of the Physics Department of the University of Tokyo 
for hospitality and support during the completion of this work. 

\end{acknowledgments}



\begin{thebibliography}{99}
\bibitem{note1}
This ordering is reminiscent of the smectic-C type structure in liquid crystals. 
Note that a smectic-A structure with $\braket{\vec{M}(\vec{r}\,)}= M_z(z) \hat{z}$ 
leads to dipole-dipole repulsion between neighboring sheets and is not favored.
\bibitem{note2}
In general, the dipole moment $\mu$ is different for terms diagonal and off-diagonal in ``spin"; 
the relevant interaction here is the transition between the two spin states, 
and thus $\mu$ is the transition spin magnetic moment. 
For example, in $^{163}$Dy the transition moment from the $\ket{11/2, 11/2}$ to the $\ket{11/2, 9/2}$ state 
is the ratio of off-diagonal to diagonal Clebsch-Gordan coefficients, 
$\sqrt{2/11}$, times the atomic magnetic moment, 10$\mu_B$, or 4.26$\mu_B$.
\end{thebibliography}
\end{document}